\documentclass[%
 reprint,
superscriptaddress,
 amsmath,amssymb,
 aps,prl
]{revtex4-2}
\usepackage{graphicx}
\usepackage{dcolumn}
\usepackage{bm}
\usepackage{hyperref,xcolor,times}

\definecolor{linkblue}{rgb}{0, 0, 1}
\hypersetup{
	colorlinks=true,       		
	linkcolor=linkblue,          	
	citecolor=linkblue,             
	urlcolor=linkblue,          
}

\begin{document}

\title{Position-correlated biphoton wavefront sensing for quantum adaptive imaging}

\author{Yi~Zheng}
\affiliation{Laboratory of Quantum Information, University of Science and Technology of China, Hefei 230026, China}
\affiliation{Anhui Province Key Laboratory of Quantum Network, University of Science and Technology of China, Hefei 230026, China}
\affiliation{CAS Center for Excellence in Quantum Information and Quantum Physics, University of Science and Technology of China, Hefei 230026, China}
	
\author{Zhao-Di~Liu}
\affiliation{Laboratory of Quantum Information, University of Science and Technology of China, Hefei 230026, China}
\affiliation{Anhui Province Key Laboratory of Quantum Network, University of Science and Technology of China, Hefei 230026, China}
\affiliation{CAS Center for Excellence in Quantum Information and Quantum Physics, University of Science and Technology of China, Hefei 230026, China}
	
\author{Jian-Shun~Tang}
\affiliation{Laboratory of Quantum Information, University of Science and Technology of China, Hefei 230026, China}
\affiliation{Anhui Province Key Laboratory of Quantum Network, University of Science and Technology of China, Hefei 230026, China}
\affiliation{CAS Center for Excellence in Quantum Information and Quantum Physics, University of Science and Technology of China, Hefei 230026, China}
\affiliation{Hefei National Laboratory, University of Science and Technology of China, Hefei 230088, China}

\author{Jin-Shi~Xu}
\email{jsxu@ustc.edu.cn}
\affiliation{Laboratory of Quantum Information, University of Science and Technology of China, Hefei 230026, China}
\affiliation{Anhui Province Key Laboratory of Quantum Network, University of Science and Technology of China, Hefei 230026, China}
\affiliation{CAS Center for Excellence in Quantum Information and Quantum Physics, University of Science and Technology of China, Hefei 230026, China}
\affiliation{Hefei National Laboratory, University of Science and Technology of China, Hefei 230088, China}

\author{Chuan-Feng~Li}
\email{cfli@ustc.edu.cn}
\affiliation{Laboratory of Quantum Information, University of Science and Technology of China, Hefei 230026, China}
\affiliation{Anhui Province Key Laboratory of Quantum Network, University of Science and Technology of China, Hefei 230026, China}
\affiliation{CAS Center for Excellence in Quantum Information and Quantum Physics, University of Science and Technology of China, Hefei 230026, China}
\affiliation{Hefei National Laboratory, University of Science and Technology of China, Hefei 230088, China}
	
\author{Guang-Can~Guo}
\affiliation{Laboratory of Quantum Information, University of Science and Technology of China, Hefei 230026, China}
\affiliation{Anhui Province Key Laboratory of Quantum Network, University of Science and Technology of China, Hefei 230026, China}
\affiliation{CAS Center for Excellence in Quantum Information and Quantum Physics, University of Science and Technology of China, Hefei 230026, China}
\affiliation{Hefei National Laboratory, University of Science and Technology of China, Hefei 230088, China}

\date{\today}

\begin{abstract}
Quantum imaging with spatially entangled photons offers advantages such as enhanced spatial resolution, robustness against noise, and counter-intuitive phenomena, while a biphoton spatial aberration generally degrades its performance. Biphoton aberration correction has been achieved by using classical beams to detect the aberration source or scanning the correction phase on biphotons if the source is unreachable. Here, a new method named position-correlated biphoton Shack--Hartmann wavefront sensing is introduced, where the phase pattern added on photon pairs with a strong position correlation is reconstructed from their position centroid distribution at the back focal plane of a microlens array. Experimentally, biphoton phase measurement and adaptive imaging against the disturbance of a plastic film are demonstrated. This single-shot method is a more direct and efficient approach toward quantum adaptive optics, suitable for integration into quantum microscopy, remote imaging, and communication.
\end{abstract}

\maketitle

\vspace{1em}
{\noindent\large{\bf Introduction}}

Entangled photons play a critical role in the development of quantum information and technology \cite{Zhang:24}. Quantum imaging, which fully utilizes the spatial degree of freedom of photons, can achieve several nonclassical optical effects \cite{Moreau2019,Defienne2024}. Ghost imaging \cite{ghostimaging} and quantum imaging with undetected photons \cite{Lemos2014} are counter-intuitive imaging methods. By joint probability distribution (JPD) measurement, spatially entangled photons can achieve a higher spatial resolution than the Rayleigh limit \cite{PhysRevA.79.013827,Toninelli:19,He2023}, which is beneficial in optical microscopy. Also, they can be distilled from the stray light \cite{distill,sciadv.aay2652}. However, the atmospheric turbulence or flaws in optical instrument introduce phase aberrations and degrade the imaging performance in both classical and quantum imaging \cite{science.adk7825}. Adaptive optics including phase measurement and correction is dedicated to overcoming this problem.

In classical optics, a famous phase measurement method is Shack--Hartmann wavefront sensing (SHWS) \cite{Shack01,Ares:00,Zheng:21}, which uses a microlens array to focus the light inside each aperture. The local obliquity of light within an aperture, corresponding to the phase gradient, is mapped to the displacement of the spot at the microlens back focal plane. The measured gradient distribution is discretized according to the microlens width, which limits its spatial resolution. Then, the phase distribution is reconstructed by the zonal or modal method \cite{Southwell:80} and corrected by a spatial light modulator (SLM) or a deformable mirror. There are also other types of wavefront sensors \cite{pyramid,Yang2020,Zheng:21} and even sensorless adaptive optical techniques. The basic idea of a notable sensorless one is that with aberration, the focused spot after a Fourier lens increases in size and decreases in peak intensity, which serves as the criterion to perform feedback control of the correction phase till the sharp peak revives \cite{Booth:06}.

In quantum optics, the spatial phase of entangled photon pairs (biphotons) can be measured by holography using a reference beam \cite{Black:23,Zia2023} or the polarization entanglement \cite{Defienne2021,sciadv.abj2155}, and the aberration cancellation of biphotons correlated in position by measuring the aberration source with classical lights has been demonstrated \cite{PhysRevLett.121.233601,sciadv.abb6298,Shekel:24} (in our article, the word ``correlated'' means the two photons are approximately at the same position, contrary to the term ``anti-correlated,'' and does not mean they can have an arbitrary nonseparable JPD). On the other hand, if a phase aberration from an unreachable source has already been added to position-correlated biphotons, a prominent measurement method has been demonstrated by Cameron \emph{et al.}\ \cite{science.adk7825}. In their protocol, inspired by the sensorless method \cite{Booth:06}, the criterion is the peak value of the biphoton position sum-coordinate (or centroid, equivalently) marginal distribution at the Fourier plane. By scanning the coefficients of Zernike polynomials, the aberration can be eliminated. This indirect way requires multiple measurement steps, so the optimal group of coefficients may not be easily obtained by sequential scanning rather than using special algorithms. In this work, we introduce a single-shot method to realize this task, namely position-correlated biphoton SHWS (PCB-SHWS), which directly measures the gradient of phase added on position-correlated biphotons. Then, using biphotons from spontaneous parametric down-conversion (SPDC) \cite{WALBORN201087,Schneeloch_2016}, the phase measurement and adaptive imaging is experimentally demonstrated.

\vspace{1em}
{\noindent\large{\bf Results}}

{\noindent\bf Theoretical framework}

The most fundamental idea of PCB-SHWS is from the Einstein--Podolsky--Rosen paper \cite{EPRpaper}. Denoting the transverse position $\boldsymbol{\rho}=(x,y)$ and momentum $\mathbf{q}=(k_x,k_y)$, letting the biphoton field in front of a lens with the focal length $f_\mathrm{SH}$ have a constant intensity, a perfect position correlation, and the phase of an oblique plane wave $e^{i\mathbf{q}_0\cdot\boldsymbol{\rho}}$ added on each photon, the wavefunction (joint amplitude) \cite{Zheng2023,Zia2023,qshws} in the position space $\psi(\boldsymbol{\rho}_1,\boldsymbol{\rho}_2)=\delta(\boldsymbol{\rho}_1-\boldsymbol{\rho}_2)e^{2i\mathbf{q}_0\cdot\boldsymbol{\rho}_1}$, and their momenta are perfectly anti-correlated $\tilde{\psi}(\mathbf{q}_1,\mathbf{q}_2)=\delta(\mathbf{q}_1+\mathbf{q}_2-2\mathbf{q}_0)$ with the anti-correlation center $\mathbf{q}_0$. Denoting their wavelength $\lambda$ and wave number $k=2\pi/\lambda$, the position wavefunction at the back focal plane of the lens is $\tilde{\psi}(f_\mathrm{SH}\mathbf{q}_1/k,f_\mathrm{SH}\mathbf{q}_2/k)$ with a paraboloid phase added [because $\psi(\boldsymbol{\rho}_1,\boldsymbol{\rho}_2)$ is not the wavefunction at the front focal plane] which does not affect the JPD $\Gamma(\mathbf{q}_1,\mathbf{q}_2)=|\tilde{\psi}(\mathbf{q}_1,\mathbf{q}_2)|^2$ (for simplicity, we directly use $\mathbf{q}$ and ignore the normalization). So, after measuring the JPD, denoting the centroid $\mathbf{q}_c=(\mathbf{q}_1+\mathbf{q}_2)/2$, summing JPD values of point pairs with the same centroid yields the biphoton centroid marginal distribution
\begin{equation}
    \Gamma_c(\mathbf{q}_c)=\int d\mathbf{q}_1\Gamma(\mathbf{q}_1,2\mathbf{q}_c-\mathbf{q}_1)\propto\delta(\mathbf{q}_c-\mathbf{q}_0)
\end{equation}
which is a sharp peak at $\mathbf{q}_0$. When photons at the whole microlens array arrive at its back focal plane, ignoring the momentum anti-correlation weakening caused by the finite microlens width $2a$, if the phase inside each aperture can be approximated by an oblique plane wave and the dynamic range of SHWS $|q_{0x}|,|q_{0y}|<ka/f_\mathrm{SH}$ is satisfied, photon pairs from different apertures have distinct centroids which are inside their own aperture, and the whole centroid distribution is an array of sharp peaks, similar as the measured intensity distribution of classical SHWS. The gradient values of the whole phase pattern can be calculated from the peak positions relative to their centers.

Then, we consider the effects of the finite microlens width, a finite position correlation with the form $c(\boldsymbol{\rho}_1-\boldsymbol{\rho}_2)$ whose width is far less than $2a$, and the added phase $\Phi(\boldsymbol{\rho})$ which is generally not a plane wave. The position wavefunction at an aperture $S$ centered by $\mathbf{0}$ is
\begin{equation}\label{poswf}
    \psi(\boldsymbol{\rho}_1,\boldsymbol{\rho}_2)=c(\boldsymbol{\rho}_1-\boldsymbol{\rho}_2)U(\boldsymbol{\rho}_1)U(\boldsymbol{\rho}_2),
\end{equation}
where $U(\boldsymbol{\rho})=e^{i\Phi(\boldsymbol{\rho})}\operatorname{rect}[\boldsymbol{\rho}/(2a)]$ and the two-dimensional rectangular function $\operatorname{rect}(\boldsymbol{\rho})=1$ when $|x|\leq1/2,|y|\leq1/2$ and $0$ elsewhere. Denoting the Fourier transforms $\tilde{U}(\mathbf{q})=\int d\boldsymbol{\rho}U(\boldsymbol{\rho})e^{-i\mathbf{q}\cdot\boldsymbol{\rho}}$ and $\tilde{c}(\mathbf{q})=\int d\boldsymbol{\rho}c(\boldsymbol{\rho})e^{-i\mathbf{q}\cdot\boldsymbol{\rho}}$, from the convolution theorem, the momentum wavefunction
\begin{equation}\label{mwfconvo}
    \tilde{\psi}(\mathbf{q}_1,\mathbf{q}_2)=[\tilde{U}(\mathbf{q}_1)\tilde{U}(\mathbf{q}_2)]\ast\left[\tilde{c}\left(\frac{\mathbf{q}_1-\mathbf{q}_2}{2}\right)\delta(\mathbf{q}_1+\mathbf{q}_2)\right],
\end{equation}
where ``$\ast$'' is the convolution sign. In classical SHWS, $|\tilde{U}(\mathbf{q})|^2$ is the measured intensity of the field $U(\boldsymbol{\rho})$, and the average $\mathbf{q}$ is the average phase gradient in this aperture \cite{Ares:00} 
\begin{equation}\label{avgqclassical}
\langle\mathbf{q}\rangle=\frac{1}{(2a)^2}\int_S d\boldsymbol{\rho}\nabla\Phi(\boldsymbol{\rho})=\langle\nabla\Phi(\boldsymbol{\rho})\rangle_S.
\end{equation}
For $\Phi(\boldsymbol{\rho})$ whose gradient is slow-varying within the aperture (in order to satisfy the spatial resolution limit), $\tilde{U}(\mathbf{q})$ is still peaked at $\langle\mathbf{q}\rangle$, although the exact distribution is a bit different from the sinc function in the plane-wave case. From Eq.\ \eqref{mwfconvo}, considering the displacement and broadening effects of convolution, their momenta are approximately anti-correlated with the center $\langle\mathbf{q}\rangle$, so the measured centroid distribution is peaked at $\langle\mathbf{q}\rangle$. The spot in the direct image is a slightly blurred $|\tilde{c}(\mathbf{q})|^2$, which may be larger than an aperture (see Supplementary Information 2 for the experimental images), as the biphoton field is spatially incoherent in the first order. So, the phase added on such a biphoton field cannot be measured by classical SHWS.

\begin{figure}[t]
\centering
\includegraphics[width=\linewidth]{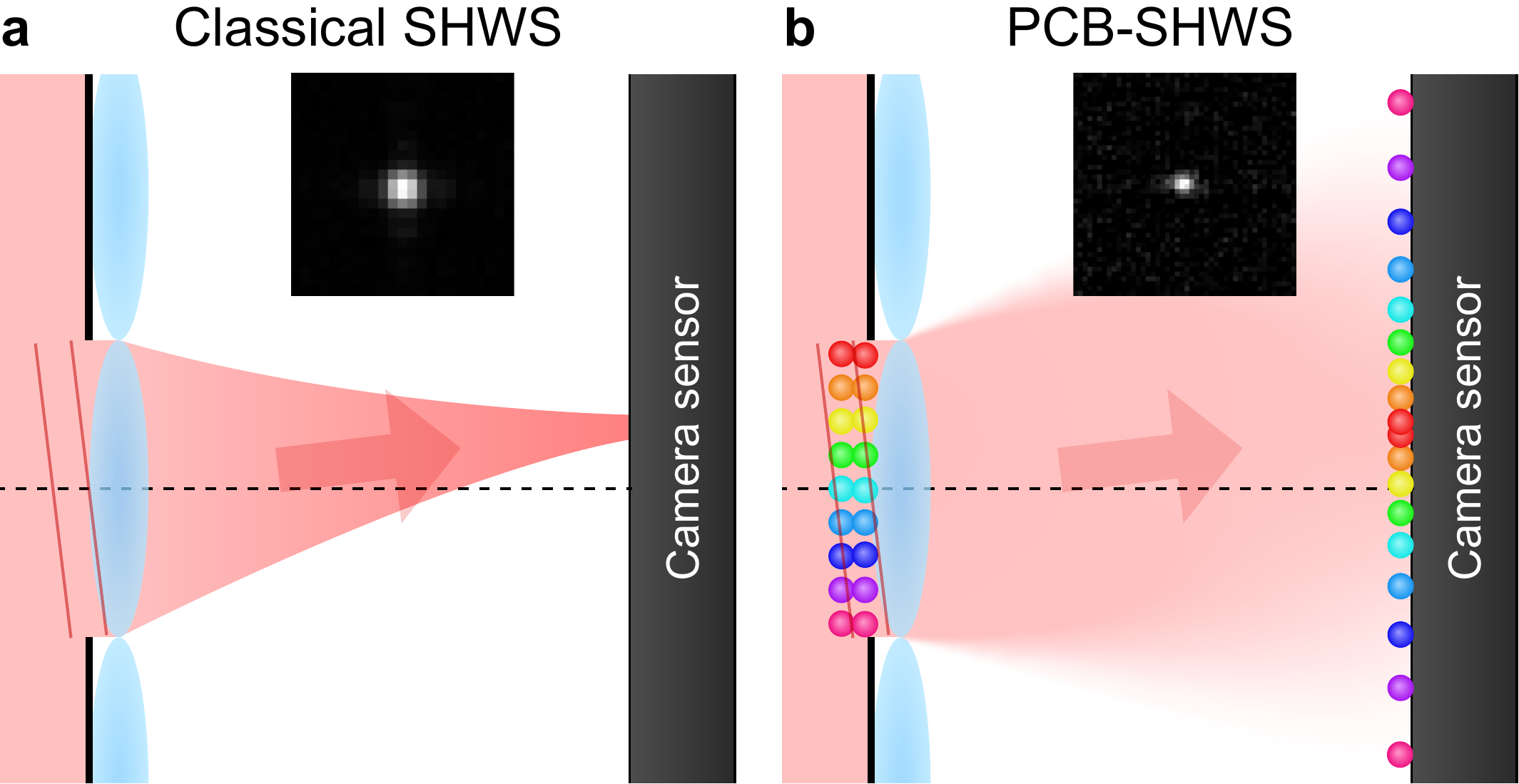}
\caption{{\bf Principle of classical SHWS and PCB-SHWS.} For simplicity, we show the light fields from a single aperture. {\bf a} Classical SHWS. The inset shows a focused spot inside a microlens aperture with the width $300$~\textmu m using a $808$-nm laser as the light source, taken by the EMCCD with the EM gain set to $0$. {\bf b} PCB-SHWS. At the microlens and its back focal plane respectively, two balls with the same color represent an entangled photon pair. A tilt phase leads to the centroid displacement. The inset is the biphoton centroid marginal distribution inside an aperture from our phase measurement experiment.}
\label{principlefig}
\end{figure}
Under the approximation of a perfect position correlation $c(\boldsymbol{\rho})\approx\delta(\boldsymbol{\rho})$, Eq.\ \eqref{mwfconvo} can be further simplified to
\begin{align}\label{mwfconvoperf}
    \tilde{\psi}(\mathbf{q}_1,\mathbf{q}_2)&\approx[\tilde{U}(\mathbf{q}_1)\tilde{U}(\mathbf{q}_2)]\ast\delta(\mathbf{q}_1+\mathbf{q}_2)\nonumber\\
    &=\int d\mathbf{q}\tilde{U}(\mathbf{q})\tilde{U}(\mathbf{q}_1+\mathbf{q}_2-\mathbf{q})\nonumber\\
    &=\tilde{U}_2(\mathbf{q}_1+\mathbf{q}_2),
\end{align}
where $\tilde{U}_2(\mathbf{q})$ is the Fourier transform of $[U(\boldsymbol{\rho})]^2$ whose phase is $2\Phi(\boldsymbol{\rho})$. The centroid distribution $\Gamma_c(\mathbf{q}_c)=|\tilde{U}_2(2\mathbf{q}_c)|^2$, so
\begin{equation}\label{avgq}
\langle\mathbf{q}_c\rangle=\frac{1}{2}\frac{1}{(2a)^2}\int_S d\boldsymbol{\rho}\nabla[2\Phi(\boldsymbol{\rho})]=\langle\nabla\Phi(\boldsymbol{\rho})\rangle_S
\end{equation}
exactly corresponds to the average phase gradient in classical SHWS, while the peak width is half the classical result. Experimentally, the centroid distribution can be efficiently measured \cite{PhysRevLett.102.253601} and has a twice pixel resolution than direct images \cite{Defienne2022super}. In spite of these differences, data processing methods of classical SHWS can be applied to the centroid distribution for phase reconstruction. The principle of classical SHWS and PCB-SHWS is shown in Fig.\ \ref{principlefig}, where the insets are a direct image of classical SHWS and a centroid distribution of PCB-SHWS inside an aperture.

\vspace{1em}
{\noindent\bf Experimental setup}

\begin{figure}[t]
\centering
\includegraphics[width=\linewidth]{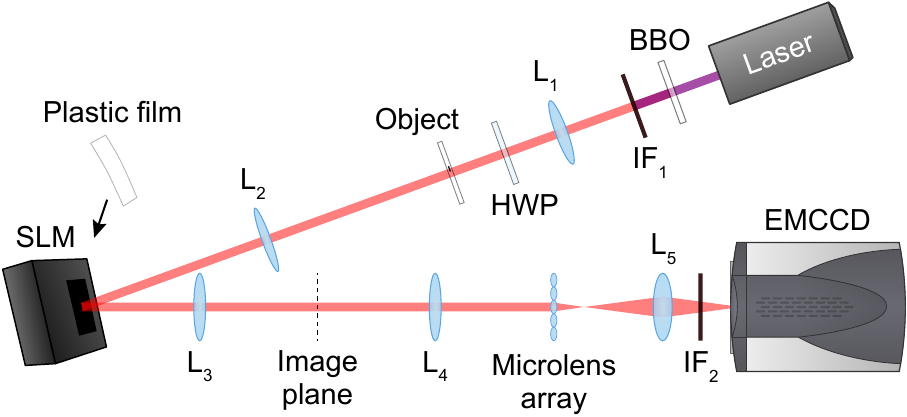}
\caption{{\bf Experimental setup of PCB-SHWS.} The laser passes through a beam shaping system (not shown) and pumps the BBO crystal. Degenerate collinear type-I down-converted photons from the BBO pass through the first Fourier lens $\mathrm{L}_1$, the half-wave plate (HWP), the object, and the second Fourier lens $\mathrm{L}_2$, and are reflected by the SLM. A plastic film may be pasted in front of the SLM. Then, they pass through a $4f$ system $\mathrm{L}_3,\mathrm{L}_4$ and is incident on the microlens array. An imaging lens $\mathrm{L}_5$ images the optical field at the microlens back focal plane to the EMCCD sensor. $\mathrm{IF}_1$: long-pass interference filter; $\mathrm{IF}_2$: bandpass filter. In the imaging experiment, the EMCCD sensor is moved to the image plane together with $\mathrm{IF}_2$ (see Supplementary Information 3 for the setup).}
\label{setupfig}
\end{figure}
We use an electron-multiplying charge-coupled device (EMCCD) camera and the multiple frame method developed by Defienne \emph{et al.}\ \cite{PhysRevLett.120.203604,PhysRevA.98.013841,Cameron_tutorial} to measure the biphoton JPD \cite{PhysRevLett.121.233601,distill,Defienne2021,sciadv.abj2155,Bhattacharjee2022,Defienne2022super,He2023,science.adk7825,qshws,Kam2025}. The setup of the phase measurement experiment is shown in Fig.\ \ref{setupfig}. A horizontally-polarized (H) laser beam at $404$~nm pumps a $\beta$-barium borate (BBO) crystal with the thickness $1$~mm and is removed by a long-pass interference filter (IF). Degenerate collinear type-I down-converted photon pairs at the vertical (V) polarization pass through the first Fourier lens ($f_1=15$~cm) and are switched to H polarization by a half-wave plate. Their positions are anti-correlated at its focal plane, where the object (USAF 1951 resolution target) is placed at one half of the beam. After the second Fourier lens ($f_2=25$~cm), an SLM adds the same phase pattern on each photon. Then, they pass through a $4f$ system ($f_3=f_4=15$~cm) and arrive at the microlens array with the aperture width $300$~\textmu m and focal length $f_\mathrm{SH}=14.6$~mm. An imaging lens ($f_5=5$~cm) projects the optical field at the microlens back focal plane to the EMCCD sensor with the magnification ratio $-1$, and a bandpass IF selects near-degenerate down-converted photons. See Materials and methods for details.

\vspace{1em}
{\noindent\bf Biphoton phase measurement}

\begin{figure}[t]
\centering
\includegraphics[width=\linewidth]{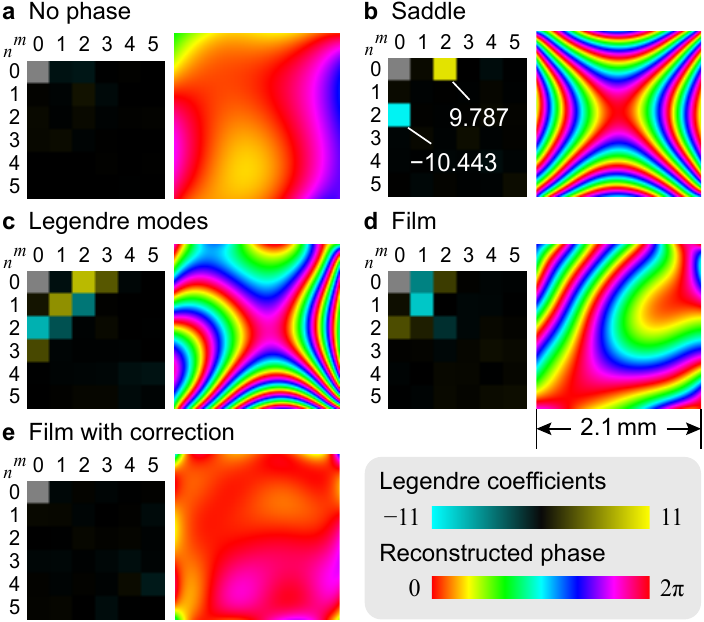}
\caption{{\bf Biphoton phase measurement result.} The calculated Legendre coefficients (excluding $L_{0,0}$) and phase distributions are shown for the five cases: {\bf a} no phase, {\bf b} saddle phase, {\bf c} Legendre modes, {\bf d} a film, and {\bf e} the film with correction. The gradient distribution of the no-phase case is the reference of the other cases.}
\label{datamainfig}
\end{figure}
\begin{figure*}[t]
\centering
\includegraphics[width=0.95\linewidth]{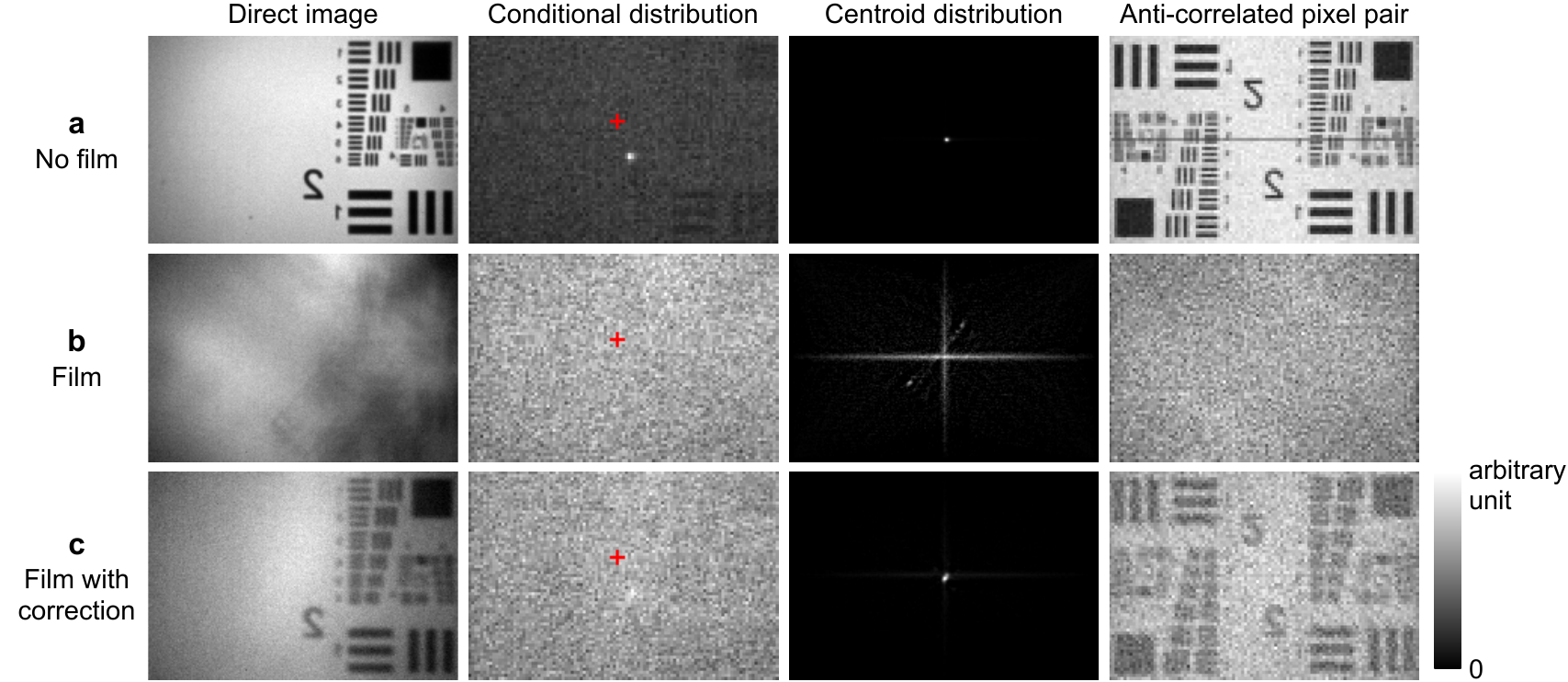}
\caption{{\bf Adaptive imaging result.} The direct images (without pixel binning, normalized according to the maximum and minimum value), CPDs with the other photon postselected to a certain pixel (the red cross), centroid marginal distributions without background removal (the horizontal and vertical bright lines are noises from the summation which are present in all three cases, but in the no-film and film with correction case, they are less significant because the peaks are much brighter), and joint probabilities of anti-correlated pixel pairs are shown for the {\bf a} no-film, {\bf b} film, and {\bf c} film with correction case. The image sizes are $2.678~\mathrm{mm}\times1.794~\mathrm{mm}$.}
\label{imagefig}
\end{figure*}
As the region of interest (ROI) in our experiment is a square, we use the modal method based on two-dimensional Legendre polynomials \cite{Niu_2022} $L_{m,n}(\boldsymbol{\rho})=L_m(x)L_n(y)$ to reconstruct the phase distribution from the gradient, rather than Zernike polynomials which are suitable for circular ROIs. See Materials and methods and Supplementary Information 1 for data processing algorithms.

We measure the phase of five cases: no phase added, a hyperbolic paraboloid (saddle) phase $10(L_{2,0}-L_{0,2})$, a superposition of several Legendre modes
\begin{equation}\label{legmodes}
    8L_{2,0}+6L_{1,1}-7L_{0,2}+4L_{3,0}-5L_{2,1}-4L_{1,2}+3L_{0,3},
\end{equation}
a plastic film placed in front of the SLM, and the plastic film with phase correction. The exposure times, direct images, centroid distributions, and gradient distributions are shown in Supplementary Information 2. From the direct images, the object pattern is faintly visible in each aperture because it modifies $\tilde{c}(\mathbf{q})$. The measured gradient distribution of the no-phase case still deviates from zero, which is mainly due to the aberration of the imaging lens, so it serves as a reference by which the measured gradient distributions of other cases are subtracted. The calculated Legendre coefficients (see the tables in Supplementary Information for values) and the reconstructed phase distributions are shown in Fig.\ \ref{datamainfig}. In the saddle phase case, the calculated $L_{2,0}$ and $L_{0,2}$ are close to the theoretical values $10$ and $-10$ respectively. In the cases of Legendre modes and film with correction, ignoring the tilt terms $L_{1,0}$ and $L_{0,1}$ which do not blur the image, compared to Eq.\ \eqref{legmodes} and $0$, respectively, the root mean square error values of the phase distributions (discretized into $120\times120$ pixels; in the unit of optical path difference) are $0.0623\lambda$ and $0.0502\lambda$. So, the measured phases fit well with the designed ones on the SLM, and the aberration caused by the film is approximately corrected. The errors are mainly from the imperfectness of SLM, slight misalignment of the setup, image distortion from the imaging lens, and noises from the multiple frame method.

\vspace{1em}
{\noindent\bf Adaptive imaging}

To test whether the biphoton phase is correctly cancelled after PCB-SHWS, we then perform the adaptive imaging experiment. The EMCCD sensor is moved to the back focal plane of the third Fourier lens together with the bandpass IF. See Supplementary Information 3 for a figure of this new setup and a brief principle of biphoton imaging with aberrations. When taking frames, $2\times2$ pixels are binned into one on the EMCCD to increase the frame rate \cite{He2023}.

We take $4.41\times10^6$ frames in each of the three cases: no film, film, and film with correction, as well as direct images without pixel binning. The direct images, conditional probability distributions (CPDs) of one photon with the other photon postselected to a certain pixel, centroid distributions, and JPDs of anti-correlated pixel pairs are shown in Fig.\ \ref{imagefig}. The anti-correlated pair distributions have central symmetry, so the object patterns exist on the other halves of the images. When the film is present, the object cannot be identified in the direct image or the anti-correlated pair distribution. After correction, they are visible again but a bit blurry because the slightly curved film about $2$~mm away from the SLM liquid crystal plate cannot be regarded as a pure phase object exactly at the SLM plane. Part of the photons are reflected or scattered after passing through the film twice, equivalent to a lower quantum efficiency of the sensor, so the signal-to-noise ratios (SNRs) of the conditional distribution and the anti-correlated pair distribution in the film with correction case are much lower than the no-film case. The centroid distribution, which is the metric of the sensorless adaptive imaging method \cite{science.adk7825}, is peaked at the center point in the no-film case or the film with correction case, while scattered in the film case.

\vspace{1em}
{\noindent\large{\bf Discussion}}

In this work, we introduced PCB-SHWS to measure the phase pattern added on biphotons with a strong position correlation, and performed experiments of Legendre coefficient measurement and adaptive imaging against the disturbance of a plastic film. Both PCB-SHWS and the sensorless approach \cite{science.adk7825} do not require classical detection of the aberration source \cite{sciadv.abb6298,Shekel:24}, reference beams \cite{Black:23}, or polarization entanglement \cite{Defienne2021,sciadv.abj2155}, while our method only requires one measurement step and has the advantage in efficiency. Compared to the multiple frame method, with a more advanced biphoton JPD measurement technique like the time-stamping camera \cite{Courme:23,Zia2023}, the phase can be rapidly acquired for real-time (truly adaptive) correction of a time-varying aberration, suitable for future quantum communication and imaging researches. However, limited by the microlens width and camera pixel width which affect the spatial resolution and sensitivity respectively, like classical SHWS, the result of PCB-SHWS may not be precise enough. So, in order to restore a quantum image from common aberrations perfectly in quantum microscopy, PCB-SHWS can first provide a good estimate of the phase, and then mode coefficient scanning can precisely determine it. By reducing the microlens size or magnifying the beam, the spatial resolution can be improved to some extent (with the sensitivity reduced), but it is after all not applicable to extremely detailed phase patterns such as those from diffusers \cite{PhysRevLett.121.233601,sciadv.abb6298,Shekel:24}.

If the photons are anti-correlated in position, the effective phase is an even function \cite{Cameron_tutorial}, and its measurement scheme can also be analyzed, but the lateral position of the microlens array should be adjusted so that the anti-correlation center is the center, a side midpoint, or a vertex of an aperture, as detailed in Supplementary Information 4.

Although the essential sensing setup (a microlens array and a camera capable of JPD measurement) in PCB-SHWS also appears in the method named quantum SHWS from our previous work \cite{qshws}, their theoretical frameworks, data processing methods, and scopes of application are distinct, and a biphoton state suitable for one method is unsuitable for the other. A clarification of their differences are given in Supplementary Information 5. Also, in classical optics, apart from using only the peak positions for SHWS, the whole momentum distribution from each microlens aperture provides more information about the optical field \cite{PhysRevLett.105.010401,Stoklasa2014}, which has been incorporated in a new technique, namely plenoptic imaging or light-field imaging \cite{PhysRevLett.116.223602,PhysRevApplied.21.024032}. Therefore, the microlens array, which enables position and local momentum measurement without breaking the uncertainty principle, has great potential in detecting higher-order correlation properties of multiphoton optical fields when combined with efficient photon coincidence detection techniques. Its further applications in quantum optics can be explored in the future.

\vspace{1em}
{\noindent\large{\bf Materials and methods}}

{\noindent\bf Details of the experimental setup}

Before pumping the BBO, the continuous-wave laser beam (TOPTICA) passes through two Fourier lenses for magnification, two cylindrical lenses to adjust its shape, and a short-pass filter at $600$~nm. The long-pass IF is at $647$~nm. The SLM (Hamamatsu X13267) has $792\times600$ pixels whose widths are $12.5$~\textmu m. We choose $f_2>f_3$ because the beam at the back focal plane of the first Fourier lens is too large for the EMCCD (Andor iXon Ultra 888; pixel width: $13$~\textmu m) in the imaging setup, and thus the image of the object is shrunk (we focus on proof-of-principle aberration cancellation in this work and do not discuss quantum superresolution imaging or image distillation). The imaging lens is used because the distance between the EMCCD casing and its sensor is $1.75$~cm, which is larger than $f_\mathrm{SH}$, and it is not easy to put the microlens array (LBTEK) inside the casing (a shutter is at its inner side). This will introduce distortions when imaging, so the beam should be at the center of the imaging lens and not too large. The center wavelength of the bandpass IF [$(810\pm5)$~nm] is not $808$~nm, but degenerate photon pairs can pass through it, and the only shortcoming is the existence of some single-photon incidences which are treated as dark counts.

Using the formulae in Ref.\ \cite{PhysRevA.110.063710}, the full width at half maximum (FWHM) of the biphoton CPD (the two-dimensional probability distribution of one photon when the other is postselected to a given position) is about $17$~\textmu m at the BBO and $28$~\textmu m at the microlens array, so most of the photon pairs pass through the same microlens; the FWHM of the spot at the back focal plane from one aperture is about $604$~\textmu m, larger than the aperture.

The ROI of the EMCCD is set to $165\times165$ pixels in the phase measurement experiment and $105\times71$ (binned) in the imaging one. The EM gain (set to $0$ when taking the direct image in Fig.\ \ref{principlefig}a), horizontal pixel readout rate, vertical pixel shift speed, and vertical clock voltage amplitude are set to $1000$, $10$~MHz, $0.6$~\textmu s, and $+2$~V respectively. The exposure times are different in each measurement, determined by the beam intensity so that roughly $1/5$ of the pixels in the ROI have photons in each frame \cite{PhysRevA.98.013841} (see Supplementary Information 2 for the values in the phase measurement experiment. In the imaging experiment, the exposure times are $1.4$~ms, $1.8$~ms, and $2.4$~ms for the no-film, film, and film with correction cases respectively). The EMCCD sensor is cooled to $-53^\circ\mathrm{C}$ and $-38^\circ\mathrm{C}$ in the phase measurement and imaging (a higher frame rate produces more heat) experiment respectively. When taking the three direct images, the exposure time is $0.1$~s, and $1000$ frames are directly summed for each case.

\vspace{1em}
{\noindent\bf Data processing}

In the multiple frame method, the EMCCD captures $N$ frames with grayscale values from $0$ to $65535$, and a threshold binarizes the frame data. In our experiment, by evaluating the grayscale value histogram without the down-converted photons, the threshold is set to $509$. Denoting the counting of the $i$th pixel (the indexing is arbitrary) of the $n$th frame as $C_{n,i}$ ($C_{n,i}=0$ or $1$), defining the single-pixel count average $\langle C_i\rangle=\frac{1}{N}\sum_{n=1}^NC_{n,i}$ and the two-pixel coincidence average $\langle C_{ij}\rangle=\frac{1}{N}\sum_{n=1}^NC_{n,i}C_{n,j}$, if the Poissonian statistics of photons is assumed, the JPD is estimated by the covariance of the counts of two pixels $\Gamma_{ij}\approx\langle C_{ij}\rangle-\langle C_i\rangle\langle C_j\rangle$. Its derivation has taken the quantum efficiency and the dark count into account \cite{PhysRevA.98.013841}. The SNR is a major issue, so it is only suitable for biphoton states with narrow CPDs \cite{Bhattacharjee2022}. The JPD of the same pixel cannot be measured by this method, and JPD of two pixels near each other on the same line has abnormal correlations due to the charge smearing effect of EMCCD \cite{Toninelli:19}, so linear interpolation is applied when necessary: for two pixels on the same line $(x_1,y)$ and $(x_2,y)$, if $|x_1-x_2|\leq10$ pixels (no such condition in the imaging experiment), their JPD value $\Gamma_{(x_1,y),(x_2,y)}$ are replaced by $(\Gamma_{(x_1,y),(x_2,y+1)}+\Gamma_{(x_1,y),(x_2,y-1)})/2$. If the photons have a strong position anti-correlation, the interpolated result will be lower than normal, as shown in the no-film case in Fig.\ \ref{imagefig}a.

The actual count rate drifts over time, adding a positive background to the calculated $\Gamma_{ij}$ \cite{PhysRevLett.120.203604}, so some works use a successive frame formula \cite{distill,Defienne2021} or discard frames with too few or too many pixels with counting \cite{qshws} to reduce it, while the SNR is reduced \cite{He2023}. In PCB-SHWS, we only need the peak positions in the centroid distribution, rather than actual values which are influenced by the background, so we use the original covariance formula to calculate the JPD. Then, the centroid distribution is calculated from the JPD, and the peak positions are extracted using an algorithm and converted to gradient values. See Supplementary Information 1 for details of data processing including the algorithms and a discussion about the SNR and the number of frames.

\vspace{1em}
{\noindent\bf Phase reconstruction}

The two-dimensional Legendre polynomials \cite{Niu_2022} $L_{m,n}(\boldsymbol{\rho})=L_m(x)L_n(y)$, where
\begin{equation}
	L_l(x)=\frac{1}{2^l}\sum_{k=0}^{\lfloor l/2\rfloor}\frac{(-1)^k(2l-2k)!}{k!(l-k)!(l-2k)!}x^{l-2k}
\end{equation}
is defined on $-1\leq x\leq1$. The gradient values are multiplied by $1.05$~mm to rescale the ROI from $2.1~\mathrm{mm}\times2.1~\mathrm{mm}$ ($7\times7$ apertures) to $2\times2$ with the dimension removed. We choose the maximum $m$ and $n$ to be $5$ (higher modes are rapidly oscillating), so there are $35$ modes excluding the constant phase mode $L_{0,0}=1$. For an aperture centered at $(x_0,y_0)$ after rescaling, denoting $L_l|_a^b=L_l(b)-L_l(a)$ and $\Lambda_l|_a^b=\int_a^bdxL_l(x)$, the equations are
\begin{gather}
    \kappa_x=\frac{1}{4a^2}\sum_{m,n}\alpha_{m,n}L_m|_{x_0-a}^{x_0+a}\Lambda_n|_{y_0-a}^{y_0+a},\nonumber\\
    \kappa_y=\frac{1}{4a^2}\sum_{m,n}\alpha_{m,n}\Lambda_m|_{x_0-a}^{x_0+a}L_n|_{y_0-a}^{y_0+a},
\end{gather}
where $2a$ is the aperture width after rescaling ($2/7$ in our experiment). Considering all $49$ apertures, we have $98$ equations. The coefficients $\alpha_{m,n}$ are solved by the least square method, and the phase is reconstructed $\Phi(\boldsymbol{\rho})=\sum_{m,n}\alpha_{m,n}L_{m,n}(\boldsymbol{\rho})$.

{\footnotesize

\vspace{1em}
{\noindent {\bf Acknowledgements}}

\noindent This work was funded by Innovation Program for Quantum Science and Technology (Grant Nos.\ 2021ZD0301200 and 2021ZD0301400), National Natural Science Foundation of China (Grant Nos.\ 92365205, 11821404, and W2411001), and USTC Major Frontier Research Program (Grant No.\ LS2030000002).

\vspace{1em}
{\noindent {\bf Author details}}

\noindent$^\textrm{1}$CAS Key Laboratory of Quantum Information, University of Science and Technology of China, Hefei 230026, China.
$^\textrm{2}$Anhui Province Key Laboratory of Quantum Network,
University of Science and Technology of China, Hefei 230026, China.
$^\textrm{3}$CAS Center for Excellence in Quantum Information and Quantum Physics,
University of Science and Technology of China, Hefei 230026, China.
$^\textrm{4}$Hefei National Laboratory, University of Science and Technology of China, Hefei 230088, China

\vspace{1em}
{\noindent {\bf Author contributions}}

\noindent Y.Z.\ conceived the idea, performed the experiments, and analyzed the data. Z.-D.L.\ and J.-S.T.\ provided assistance on data acquisition and processing. J.-S.X., C.-F.L., and G.-C.G.\ supervised the project. All authors participated in the discussion and analysis of the manuscript.

\vspace{1em}
{\noindent {\bf Data availability}}

\noindent All data needed to evaluate the conclusions in this paper are present in the paper and/or the Supplementary Information. Further data, including raw camera frames, are available from the corresponding author upon reasonable request.

\vspace{1em}
{\noindent {\bf Conflict of interest}}

\noindent The authors declare no competing interests.

\vspace{1em}
{\noindent \hyperlink{smpage}{\bf Supplementary information}}
}

\clearpage
\newpage
\linespread{1.16}
\setcounter{page}{1}
\appendix
\setcounter{equation}{0}
\setcounter{figure}{0}
\setcounter{table}{0}
\renewcommand{\thefigure}{S\arabic{figure}}
\renewcommand{\theequation}{S\arabic{equation}}
\renewcommand{\thetable}{S\arabic{table}}
\onecolumngrid
\renewcommand{\appendixname}{Section}

\begin{center}
    \hypertarget{smpage}{\textbf{\Large{Supplementary information for}}}

    \textbf{\large{Position-correlated biphoton wavefront sensing for quantum adaptive imaging}}
    
    \vspace{5mm}
    {\small Yi~Zheng, Zhao-Di~Liu, Jian-Shun~Tang, Jin-Shi~Xu, Chuan-Feng~Li, and Guang-Can~Guo}
\end{center}

\vspace{1em}
{\large{\bf 1. Details of data processing}}
\vspace{0.2em}

When calculating the centroid distribution, if joint probabilities of all pixel pairs with the same centroid position are summed (even if they are far from each other and thus cannot have signals), the noise increases, so we choose to do the truncation: for two pixels $(x_1,y_1)$ and $(x_2,y_2)$, if $|x_1-x_2|$ or $|y_1-y_2|$ is greater than $300$~\textmu m, they do not contribute to the summing.

\begin{figure}[b]
\centering
\includegraphics[width=\linewidth]{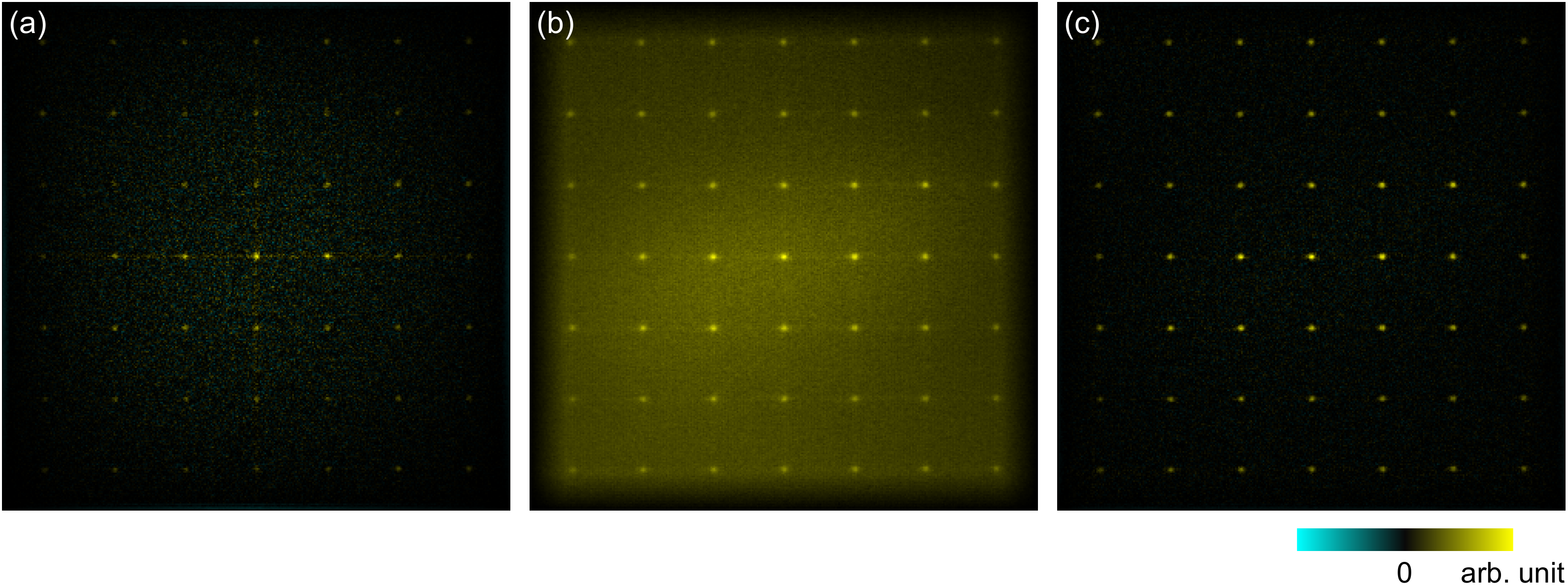}
\caption{The calculated centroid marginal distribution from the no-phase data (a) without truncation, with background removal; (b) with truncation, without background removal; (c) with truncation and background removal, which is to be used to extract the gradients.}
\label{trunc300fig}
\end{figure}

Before finding the peak positions, the background in the centroid distribution (due to the EMCCD count rate drift) is removed by this method: the value of each pixel is subtracted by the median of at most $9\times9$ pixels centered by this pixel. Figure \ref{trunc300fig} shows the distribution to be used and the cases without truncation or background removal.

\begin{figure}[b!]
\centering
\includegraphics[width=.75\linewidth]{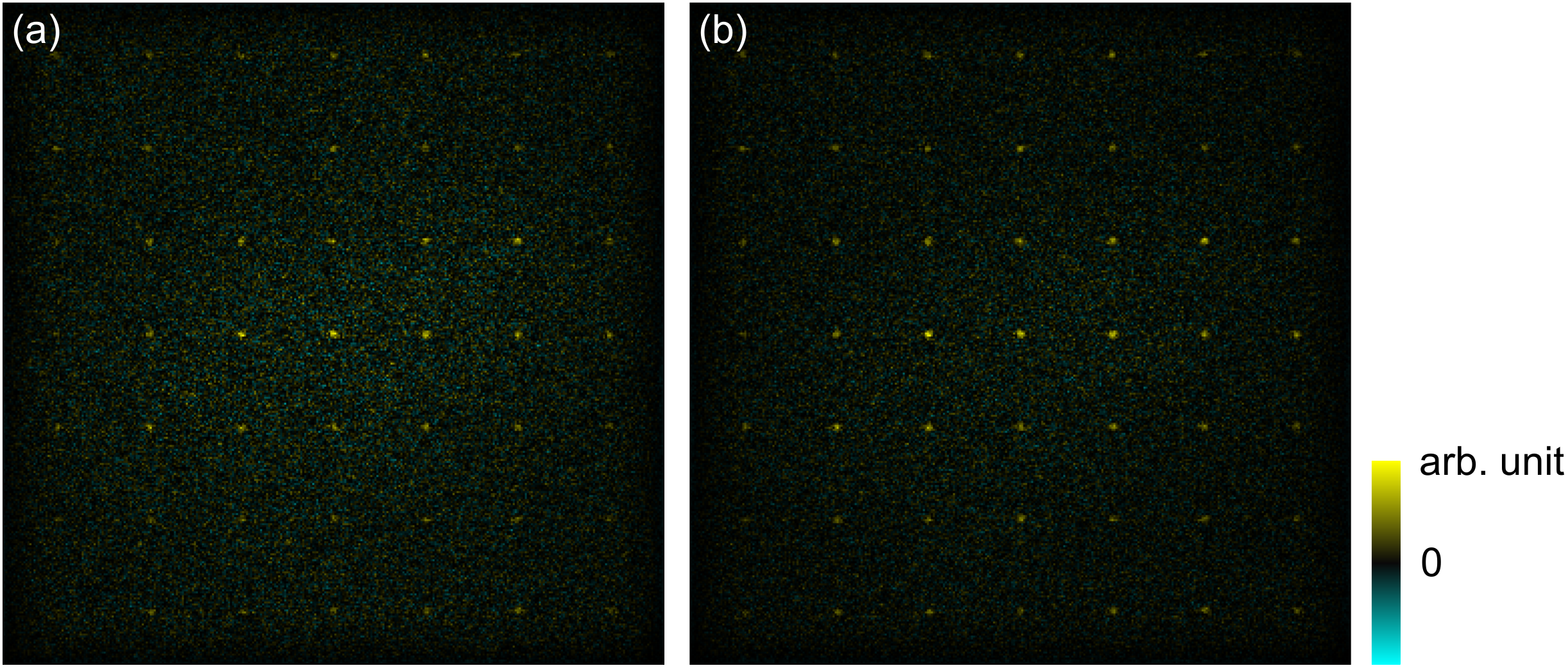}
\caption{The calculated centroid marginal distribution from the no-phase data with truncation using (a) $2\times10^4$ frames; (b) $4\times10^4$ frames.}
\label{sup24fig}
\end{figure}

\begin{figure}[t!]
\centering
\includegraphics[width=.45\linewidth]{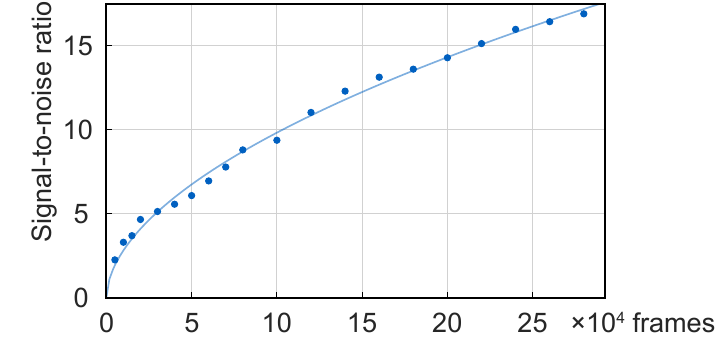}
\caption{The relation of the signal-to-noise ratio and number of frames.}
\label{supsnrfig}
\end{figure}

In our phase measurement experiment, $N$ ranges from $2.8\times10^5$ to $3.6\times10^5$. Figure \ref{sup24fig} shows the centroid distributions (with truncation and background removal) from the no-phase data using only $2\times10^4$ and $4\times10^4$ frames. From the centroid distributions inside the center aperture from various numbers of frames $N$ ranging from $5\times10^3$ to $2.8\times10^5$, we calculate their SNRs. The signal value is taken to be the average value of the $2\times2$-pixel region inside the center aperture with the largest value sum, the noise value is the standard deviation of the values in the center aperture excluding the $10\times10$-pixel region centered by the aforementioned $2\times2$-pixel region, and the SNRs are calculated and fitted by the power function $0.01888N^{0.5431}$ ($R^2\approx0.9944$), as shown in Fig.\ \ref{supsnrfig}. So, the experiment could be faster if one would like to quickly estimate the phase. The frame rate of the EMCCD with the ROI set to $165\times165$ pixels is about $37$ frames per second, so it only takes about $20$ minutes for the phase to be properly measured.

In each aperture of the centroid distribution after background removal, the peak position is calculated by first finding a $4\times4$-pixel (with half the EMCCD pixel width $6.5$~\textmu m) region with the largest value sum and then calculating the average position in the $8\times8$-pixel region centered by the $4\times4$-pixel region weighted by the value of the distribution, so that the influence of areas without the peak is reduced. Under the paraxial approximation, the phase gradient of the aperture $\boldsymbol{\kappa}$ is the average position relative to the aperture center multiplied by $k/f_\mathrm{SH}$.

\vspace{2em}
{\large{\bf 2. Data of the phase measurement experiment}}
\vspace{0.2em}

\begin{figure}[h]
\centering
\includegraphics[width=0.9\linewidth]{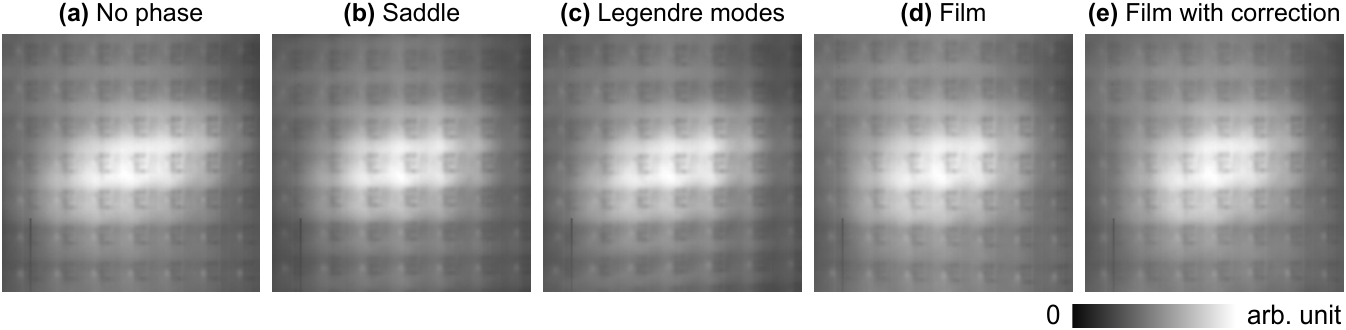}
\caption{The direct images by summing all the thresholded frames of the five cases.}
\label{supdirectfig}
\end{figure}

Figure \ref{supdirectfig} shows the direct images of the five cases in the phase measurement experiment. By comparing the no-phase case with the other images, displacements of the faint object patterns (corresponding to the phase gradient) can be observed. Certainly, this method has a lower accuracy and cannot succeed when the object is not present.

\begin{figure}[h]
\centering
\includegraphics[width=\linewidth]{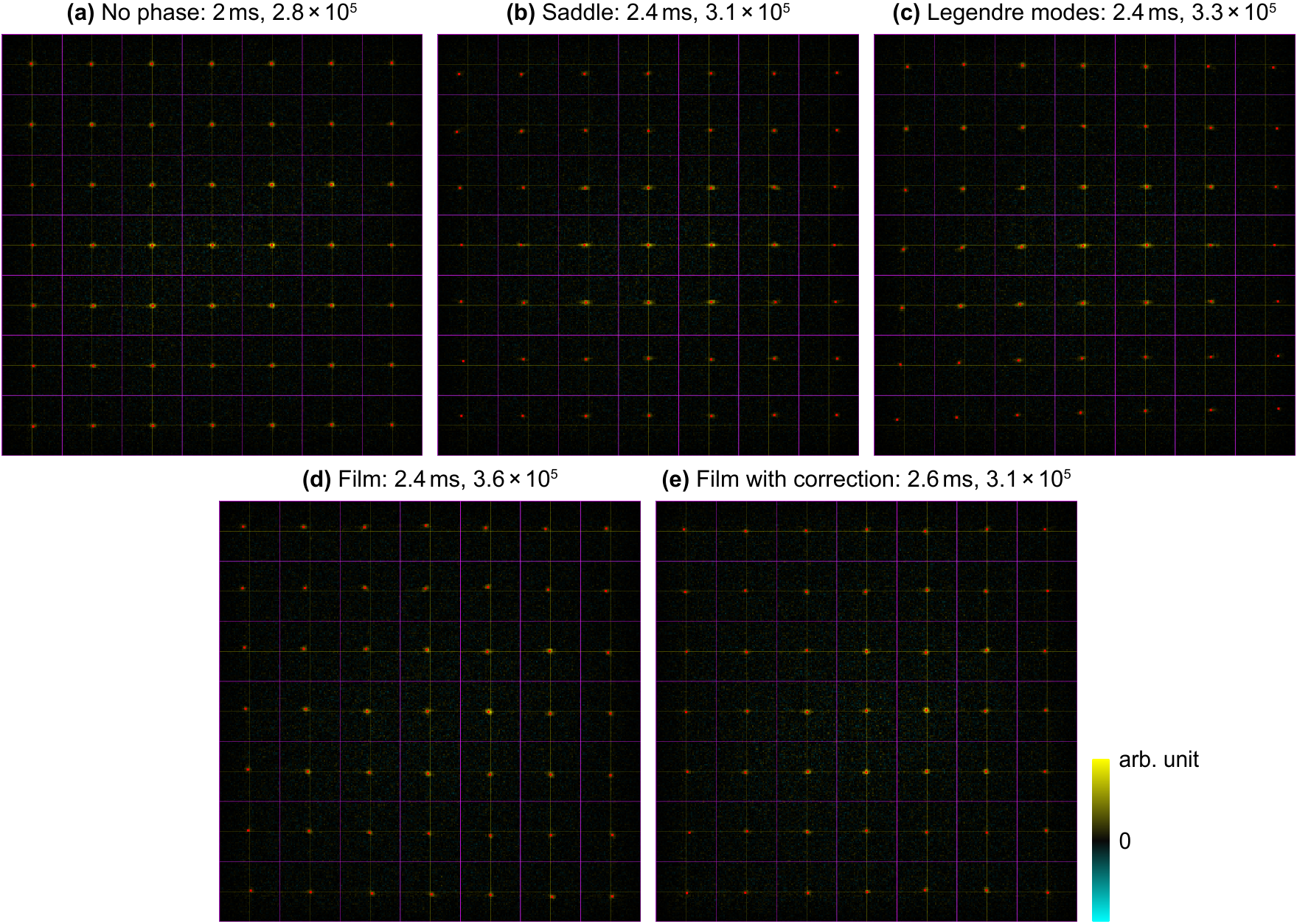}
\caption{The centroid distributions. Purple grids are aperture borders, yellow grid points are aperture centers, and red dots are the peaks identified by our algorithm. The EMCCD exposure time and number of frames taken are given for each case.}
\label{supcentroidfig}
\end{figure}

\begin{figure}[h!]
\centering
\includegraphics[width=0.56\linewidth]{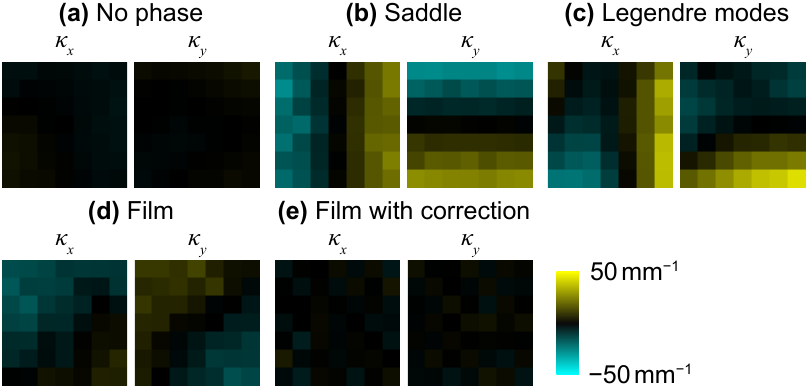}
\caption{The gradient $(\kappa_x,\kappa_y)$ distributions. The no-phase case is the reference for all other cases.}
\label{supgradfig}
\end{figure}

Figure \ref{supcentroidfig} shows the centroid distributions of the five cases with background removal and the peak positions identified by our algorithm, together with the exposure times and numbers of frames. The gradient distributions are shown in Fig.\ \ref{supgradfig}. The calculated Legendre coefficients (rounded to three decimal places) are shown in Table \ref{tablenophase}--\ref{tablefilmcor} in the last page.

\vspace{2em}
{\large{\bf 3. Experimental setup and principle of adaptive imaging}}
\vspace{0.2em}

\begin{figure}[h!]
\centering
\includegraphics[width=0.5\linewidth]{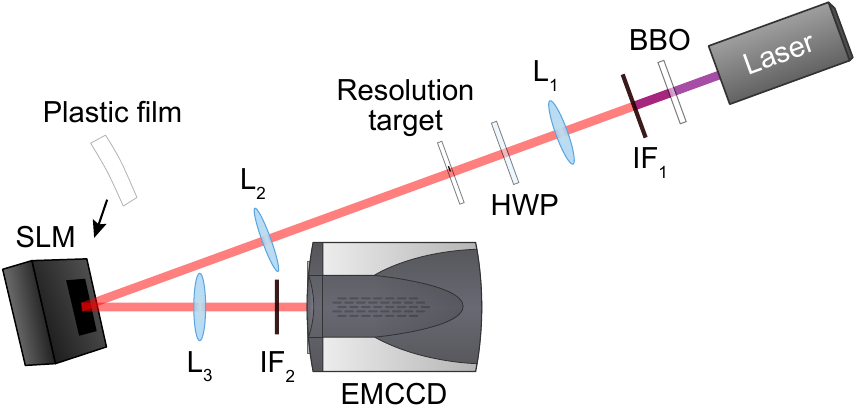}
\caption{The setup of the adaptive imaging experiment. The EMCCD sensor is at the back focal plane of $\mathrm{L}_3$.}
\label{setupimagingfig}
\end{figure}

The setup of our adaptive imaging experiment is shown in Fig.\ \ref{setupimagingfig}.

To illustrate the impact of biphoton aberration on imaging based on the setup, for simplicity, we ignore the focal length differences or finite sizes of the Fourier lenses, let the pump beam be a plane wave, and use the momentum wavefunction to describe the biphoton field after an odd number of lenses. Letting the position wavefunction at the nonlinear crystal be $\psi_0(\boldsymbol{\rho}_1,\boldsymbol{\rho}_2)=c(\boldsymbol{\rho}_1-\boldsymbol{\rho}_2)$, at the back focal plane of the first Fourier lens where the object $T(\mathbf{q})$ is placed, the wavefunction becomes $\tilde{\psi}_1(\mathbf{q}_1,\mathbf{q}_2)=\tilde{c}(\mathbf{q}_1)T(\mathbf{q}_1)T(-\mathbf{q}_1)\delta(\mathbf{q}_1+\mathbf{q}_2)$, where $\tilde{c}(\mathbf{q})$ is the Fourier transform of $c(\boldsymbol{\rho})$ [for collinear type-I SPDC, it has the form $\operatorname{sinc}(a|\mathbf{q}|^2)$, so we let $c(\boldsymbol{\rho})$ and $\tilde{c}(\mathbf{q})$ be even functions]. Denoting the effective image amplitude as $\tilde{c}'(\mathbf{q})=\tilde{c}(\mathbf{q})T(\mathbf{q})T(-\mathbf{q})$ (also even), then $\tilde{\psi}_1(\mathbf{q}_1,\mathbf{q}_2)=\tilde{c}'(\mathbf{q}_1)\delta(\mathbf{q}_1+\mathbf{q}_2)$. If the aberration source is regarded as a thin object, letting its amplitude transmittance be $T_a(\boldsymbol{\rho})$ whose Fourier transform is $h(\mathbf{q})$, the wavefunction at the imaging plane
\begin{equation}
    \tilde{\psi}_3(\mathbf{q}_1,\mathbf{q}_2)=[\tilde{c}'(\mathbf{q}_1)\delta(\mathbf{q}_1+\mathbf{q}_2)]\ast[h(\mathbf{q}_1)h(\mathbf{q}_2)].
\end{equation}
With no amplitude or phase modulation, $T_a(\boldsymbol{\rho})=1$, $h(\mathbf{q})=\delta(\mathbf{q})$, and $\tilde{\psi}_3(\mathbf{q}_1,\mathbf{q}_2)=\tilde{\psi}_1(\mathbf{q}_1,\mathbf{q}_2)$, which means the effective image $|\tilde{c}'(\mathbf{q})|^2$ can be precisely obtained in both the direct image and the JPD of anti-correlated position pairs $|\tilde{\psi}_3(\mathbf{q},-\mathbf{q})|^2$. If $T_a(\boldsymbol{\rho})$ is limited by an aperture, $h(\mathbf{q})$ and the centroid distribution have finite widths and the measured image is slightly blurry. Detecting the JPD of anti-correlated position pairs can realize quantum superresolution imaging at the standard quantum limit$^{\textrm{\hyperlink{ref1}{1}}}$.

With a phase aberration $\arg T_a(\boldsymbol{\rho})$, $h(\mathbf{q})$ is generally wide. The wavefunction of anti-correlated positions
\begin{equation}
    \tilde{\psi}_3(\mathbf{q},-\mathbf{q})=\int d\mathbf{q}'_1d\mathbf{q}'_2\tilde{c}'(\mathbf{q}'_1)\delta(\mathbf{q}'_1+\mathbf{q}'_2)h(\mathbf{q}-\mathbf{q}'_1)h(-\mathbf{q}-\mathbf{q}'_2)=\int d\mathbf{q}'\tilde{c}'(\mathbf{q}')h(\mathbf{q}-\mathbf{q}')h(-\mathbf{q}+\mathbf{q}')=\tilde{c}'(\mathbf{q})\ast h_2(\mathbf{q}),
\end{equation}
where $h_2(\mathbf{q})=h(\mathbf{q})h(-\mathbf{q})$ is also wide. So, the JPD of anti-correlated positions $|\tilde{\psi}_3(\mathbf{q},-\mathbf{q})|^2$ is blurry and the object cannot be observed.

As for the direct image, one can analyze it classically. At the object plane, the light field with the intensity $|\tilde{c}(\mathbf{q})T(\mathbf{q})|^2$ is spatially incoherent in the first order. So, at the imaging plane, the intensity is $|\tilde{c}(\mathbf{q})T(\mathbf{q})|^2\ast|h(\mathbf{q})|^2$, also vague.

\vspace{2em}
{\large{\bf 4. Theory of phase measurement of anti-correlated biphotons}}
\vspace{0.2em}

We consider photon pairs perfectly anti-correlated in position and with an infinite size $\int d\boldsymbol{\rho}|\boldsymbol{\rho}\rangle|-\boldsymbol{\rho}\rangle$. The anti-correlation center is unlikely to be the center, a side midpoint, or a vertex of an aperture, and thus for photons within a certain aperture, their entangled partners generally do not lie inside a single aperture, but at most $4$ apertures. So, the microlens array should be laterally displaced to satisfy the requirement. However, effective methods to judge whether it is displaced correctly needs to be discovered before an actual experiment.

If a phase object $\Phi(\boldsymbol{\rho})$ is added to the photon pairs, the phase term of the resulting state
\begin{equation}
    \int d\boldsymbol{\rho}e^{i[\Phi(\boldsymbol{\rho})+\Phi(-\boldsymbol{\rho})]}|\boldsymbol{\rho}\rangle|-\boldsymbol{\rho}\rangle
\end{equation}
is an even function which we denote by $\Phi_\mathrm{eff}(\boldsymbol{\rho})=\Phi(\boldsymbol{\rho})+\Phi(-\boldsymbol{\rho})$, which cancels if $\Phi(\boldsymbol{\rho})$ is an odd function up to a constant $\Phi(-\boldsymbol{\rho})=C-\Phi(\boldsymbol{\rho})$. We consider photon 1 in an aperture with the width $2a$ centered by $\boldsymbol{\rho}_0$, and photon 2 in the opposite aperture centered by $-\boldsymbol{\rho}_0$ (temporarily masking other apertures). For Fourier transform, we displace the coordinate origin of photon 1 by $\boldsymbol{\rho}_0$, and photon 2 by $-\boldsymbol{\rho}_0$. Denoting the square region with the width $2a$ centered by $\mathbf{0}$ (the origin of the coordinate system for photon 1 or 2) as $S$, the state becomes $\int_S d\boldsymbol{\rho}e^{i\Phi_\mathrm{eff}(\boldsymbol{\rho}+\boldsymbol{\rho}_0)}|\boldsymbol{\rho}\rangle|-\boldsymbol{\rho}\rangle$. Denoting $U(\boldsymbol{\rho})=e^{i\Phi_\mathrm{eff}(\boldsymbol{\rho}+\boldsymbol{\rho}_0)}\operatorname{rect}[\boldsymbol{\rho}/(2a)]$ and $\tilde{U}(\mathbf{q})=\int d\boldsymbol{\rho}U(\boldsymbol{\rho})e^{-i\mathbf{q}\cdot\boldsymbol{\rho}}$, the momentum wavefunction $\tilde{\psi}(\mathbf{q}_1,\mathbf{q}_2)=\tilde{U}(\mathbf{q}_1-\mathbf{q}_2)$. When $\boldsymbol{\rho}$ is inside $S$, considering the spatial resolution requirement of SHWS, approximate $\Phi_\mathrm{eff}(\boldsymbol{\rho}+\boldsymbol{\rho}_0)$ by $2\mathbf{q}_0\cdot\boldsymbol{\rho}$ (ignoring the constant phase), and then $\tilde{\psi}(\mathbf{q}_1,\mathbf{q}_2)=\operatorname{sinc}_2[a(\mathbf{q}_1-\mathbf{q}_2-2\mathbf{q}_0)]$, where $\operatorname{sinc}_2(\mathbf{q})=\operatorname{sinc}(q_x)\operatorname{sinc}(q_y)$. If the two aperture are the same at the center ($\boldsymbol{\rho}_0=\mathbf{0}$), $\mathbf{q}_0$ must be $\mathbf{0}$ from the even nature of $\Phi_\mathrm{eff}(\boldsymbol{\rho})$. If not, by measuring the biphoton position difference $f_\mathrm{SH}(\mathbf{q}_1-\mathbf{q}_2)/k$ marginal distribution at the microlens focal plane, $\mathbf{q}_0$ can be determined from the difference between the peak position and the displacement from the photon 2 aperture to the photon 1 aperture. When the dynamic range requirement of SHWS is satisfied, there is no crosstalk from different pairs of apertures opposite to each other, so the phase gradient distribution, which should be an odd function, can be measured from the whole position difference distribution.

\vspace{2em}
{\large{\bf 5. Difference between PCB-SHWS and quantum SHWS}}
\vspace{0.2em}

In our previous work, the method named quantum SHWS$^{\textrm{\hyperlink{ref2}{2}}}$ (we suggest another name ``SHWS of biphoton joint phase'' for it, since PCB-SHWS also involves the quantum optical field) was designed to reconstruct the phase of the four-dimensional biphoton spatial wavefunction $\psi(\boldsymbol{\rho}_1,\boldsymbol{\rho}_2)$. Its basic data processing methods are as follows. Letting the ``aperture CPD'' of one aperture be the sum of the CPDs (from the measured JPD at the back focal plane) with the other photon postselected to all pixels inside this aperture, it is used to extract the phase gradient distribution of the conditional wavefunction with the other photon postselected to this aperture which is approximated by a point. The dynamic range requirement of quantum SHWS is that photons passing through one aperture at the microlens must not escape to other apertures at the back focal plane, otherwise the calculated aperture CPD contains the contributions of the cases where the postselected photon is from other apertures. This can be judged from the camera direct image. If the larger spots (the biphoton field usually has a low spatial coherence in the first order) is still confined by their own apertures, the dynamic range requirement is satisfied, while Fig.\ \ref{supdirectfig} is a counterexample. After obtaining the phase gradient distributions of all apertures in the ROI, the four-dimensional phase can be reconstructed. Because the aperture needs approximation, the JPD and the phase gradient distribution of the biphoton state to be measured (right before the microlens array) should not vary rapidly at the scale of the microlens width, which is the spatial resolution limit of quantum SHWS. So, it only suits for biphoton states with much wider CPDs at the microlens array (e.g., with a weak or no position correlation or anti-correlation). Apart from the dynamic range and the spatial resolution, there are no other assumptions on the wavefunction. On the contrary, the theory of PCB-SHWS has assumed the biphoton has a strong position correlation, so the wavefunction of a biphoton state suitable for PCB-SHWS cannot be measured by quantum SHWS.

Since the position correlation (or the CPD width) of actual biphoton states must be finite, after magnifying the optical field many times using $4f$ systems, the resulting state can satisfy the spatial resolution requirement (the CPD width should be at least twice the microlens width) and seemingly applies to quantum SHWS. However, the whole beam is too large to be taken by a camera, and, most importantly, the much reduced phase gradients after magnification are imperceptible. So, if the biphoton state is known to have a strong position correlation, with an unknown phase added, the use of quantum SHWS is impractical.

The basic data processing method of PCB-SHWS is to calculate the biphoton centroid marginal distribution (which is two-dimensional) from the measured four-dimensional JPD. As the form of the biphoton state has been assumed, from the theory in the main text, the centroid distribution is an array of sharp peaks, which are then processed in the same way as classical SHWS to reconstruct the two-dimensional phase $\Phi(\boldsymbol{\rho})$. The dynamic range and the spatial resolution limit the form of $\Phi(\boldsymbol{\rho})$, rather than the four-dimensional wavefunction. A photon pair whose state is suitable for quantum SHWS are likely to go through different microlenses, so their centroid distribution at the back focal plane generally has no special properties.

Since the preparation of an arbitrary biphoton spatial state is difficult, we used quantum SHWS in a proof-of-principle experiment to reconstruct the biphoton wavefunction from SPDC after free-space propagation, where the weakening of the position correlation and the emergence of phase correlation had been predicted theoretically$^{\textrm{\hyperlink{ref3}{3}}}$, while real applications of quantum SHWS rely on more mature multiphoton nonlinear modulation techniques or the realization of linear optical systems with desired impulse response functions$^{\textrm{\hyperlink{ref4}{4}}}$. Also, in our quantum SHWS experiment, millions of frames were taken for a desirable SNR, which took several hours for each biphoton state even using the faster Andor iXon Ultra 897 camera. For PCB-SHWS, the two-dimensional phase modulation of position-correlated biphotons is easy (by using an SLM), and the cancellation of an existing phase aberration can be checked by imaging$^{\textrm{\hyperlink{ref5}{5}}}$. Since the centroid distribution is used, taking tens of thousands of frames is enough. So, PCB-SHWS can be more easily applied in current quantum imaging techniques.

\vspace{2em}
{\large{\bf References}}
\vspace{0.2em}

\hypertarget{ref1}1.\;Giovannetti, V., Lloyd, S., Maccone, L. \& Shapiro, J.\ H.
Sub-Rayleigh-diffraction-bound quantum imaging.
\emph{Phys.\ Rev.\ A}\ {\bf 79}, 013827 (2009).

\hypertarget{ref2}2.\;Zheng, Y.\ et al.
Characterizing biphoton spatial wave function dynamics with quantum wavefront sensing.
\emph{Phys.\ Rev.\ Lett.}\ {\bf 133}, 033602 (2024).

\hypertarget{ref3}3.\;Chan, K.\ W., Torres, J.\ P.\ \& Eberly, J.\ H.
Transverse entanglement migration in Hilbert space.
\emph{Phys.\ Rev.\ A}\ {\bf 75}, 050101(R) (2007).

\hypertarget{ref4}4.\;Zheng, Y., Xu, J.-S., Li, C.-F. \& Guo, G.-C.
Theory of the monochromatic advanced-wave picture and applications in biphoton optics.
\emph{Phys.\ Rev.\ A}\ {\bf 110}, 063710 (2024).

\hypertarget{ref5}5.\;Cameron, P.\ et al.
Adaptive optical imaging with entangled photons,
\emph{Science}\ {\bf 383}, 1142 (2024).

\clearpage
\newpage

\begin{table}[h!]
\centering
\caption{Legendre coefficients of the no-phase case}
\begin{tabular}{crrrrrr}
\hline
$m$ & \multicolumn{1}{c}{$0$} & \multicolumn{1}{c}{$1$} & \multicolumn{1}{c}{$2$} & \multicolumn{1}{c}{$3$} & \multicolumn{1}{c}{$4$} & \multicolumn{1}{c}{$5$} \\
\hline
$n=0$ & \multicolumn{1}{c}{N/A} & $-0.790$ & $-0.972$ & $-0.079$ & $0.096$ & $0.032$ \\
$n=1$ & $0.158$ & $-0.172$ & $0.776$ & $-0.424$ & $-0.027$ & $-0.041$ \\
$n=2$ & $0.425$ & $-0.124$ & $0.396$ & $-0.142$ & $0.100$ & $0.139$ \\
$n=3$ & $0.356$ & $0.489$ & $-0.178$ & $0.064$ & $-0.013$ & $0.013$ \\
$n=4$ & $0.004$ & $0.006$ & $0.039$ & $0.100$ & $-0.009$ & $0.053$ \\
$n=5$ & $0.001$ & $-0.042$ & $0.030$ & $-0.003$ & $-0.054$ & $-0.011$ \\
\hline
\end{tabular}
  \label{tablenophase}
\end{table}
\begin{table}[h!]
\centering
\caption{Legendre coefficients of the saddle phase case}
\begin{tabular}{crrrrrr}
\hline
$m$ & \multicolumn{1}{c}{$0$} & \multicolumn{1}{c}{$1$} & \multicolumn{1}{c}{$2$} & \multicolumn{1}{c}{$3$} & \multicolumn{1}{c}{$4$} & \multicolumn{1}{c}{$5$} \\
\hline
$n=0$ & \multicolumn{1}{c}{N/A} & $0.450$ & $9.787$ & $0.031$ & $-0.538$ & $-0.065$ \\
$n=1$ & $0.438$ & $-0.167$ & $0.053$ & $-0.087$ & $0.272$ & $-0.119$ \\
$n=2$ & $-10.443$ & $0.093$ & $-0.124$ & $0.127$ & $0.412$ & $0.201$ \\
$n=3$ & $-0.249$ & $0.385$ & $-0.138$ & $0.595$ & $-0.164$ & $0.146$ \\
$n=4$ & $-0.174$ & $-0.044$ & $-0.282$ & $-0.206$ & $-0.547$ & $-0.174$ \\
$n=5$ & $0.064$ & $-0.312$ & $0.185$ & $-0.094$ & $-0.170$ & $0.557$ \\
\hline
\end{tabular}
  \label{tablesaddle}
\end{table}
\begin{table}[h!]
\centering
\caption{Legendre coefficients of the Legendre modes case}
\begin{tabular}{crrrrrr}
\hline
$m$ & \multicolumn{1}{c}{$0$} & \multicolumn{1}{c}{$1$} & \multicolumn{1}{c}{$2$} & \multicolumn{1}{c}{$3$} & \multicolumn{1}{c}{$4$} & \multicolumn{1}{c}{$5$} \\
\hline
$n=0$ & \multicolumn{1}{c}{N/A} & $-0.655$ & $8.034$ & $3.748$ & $-0.296$ & $-0.039$ \\
$n=1$ & $0.881$ & $6.203$ & $-5.111$ & $0.099$ & $-0.026$ & $-0.006$ \\
$n=2$ & $-7.633$ & $-3.508$ & $-0.142$ & $0.412$ & $-0.004$ & $-0.076$ \\
$n=3$ & $3.102$ & $-0.220$ & $-0.129$ & $-0.115$ & $-0.144$ & $0.017$ \\
$n=4$ & $-0.061$ & $-0.206$ & $-0.248$ & $-0.278$ & $-0.651$ & $-0.809$ \\
$n=5$ & $-0.041$ & $0.097$ & $0.280$ & $0.283$ & $-0.442$ & $-0.187$ \\
\hline
\end{tabular}
  \label{tablelegen}
\end{table}
\begin{table}[h!]
\centering
\caption{Legendre coefficients of the film case}
\begin{tabular}{crrrrrr}
\hline
$m$ & \multicolumn{1}{c}{$0$} & \multicolumn{1}{c}{$1$} & \multicolumn{1}{c}{$2$} & \multicolumn{1}{c}{$3$} & \multicolumn{1}{c}{$4$} & \multicolumn{1}{c}{$5$} \\
\hline
$n=0$ & \multicolumn{1}{c}{N/A} & $-5.606$ & $2.581$ & $0.089$ & $-0.247$ & $-0.164$ \\
$n=1$ & $0.603$ & $-8.585$ & $0.003$ & $-0.301$ & $-0.288$ & $-0.043$ \\
$n=2$ & $3.321$ & $1.312$ & $-2.029$ & $-0.343$ & $0.345$ & $0.202$ \\
$n=3$ & $0.122$ & $0.174$ & $-0.342$ & $0.260$ & $0.563$ & $0.391$ \\
$n=4$ & $-0.204$ & $0.017$ & $0.349$ & $0.327$ & $-0.044$ & $0.300$ \\
$n=5$ & $-0.047$ & $-0.179$ & $0.397$ & $0.242$ & $0.475$ & $0.595$ \\
\hline
\end{tabular}
  \label{tablefilm}
\end{table}
\begin{table}[h!]
\centering
\caption{Legendre coefficients of the film with correction case}
\begin{tabular}{crrrrrr}
\hline
$m$ & \multicolumn{1}{c}{$0$} & \multicolumn{1}{c}{$1$} & \multicolumn{1}{c}{$2$} & \multicolumn{1}{c}{$3$} & \multicolumn{1}{c}{$4$} & \multicolumn{1}{c}{$5$} \\
\hline
$n=0$ & \multicolumn{1}{c}{N/A} & $-0.329$ & $0.198$ & $-0.227$ & $-0.077$ & $0.111$ \\
$n=1$ & $0.368$ & $0.320$ & $-0.252$ & $0.001$ & $0.123$ & $-0.450$ \\
$n=2$ & $0.022$ & $-0.046$ & $0.322$ & $0.136$ & $0.243$ & $-0.168$ \\
$n=3$ & $-0.282$ & $-0.319$ & $-0.101$ & $-0.314$ & $-0.585$ & $-0.190$ \\
$n=4$ & $0.160$ & $-0.178$ & $0.141$ & $-0.145$ & $0.484$ & $-1.109$ \\
$n=5$ & $0.179$ & $0.265$ & $-0.039$ & $-0.546$ & $-0.075$ & $-0.005$ \\
\hline
\end{tabular}
  \label{tablefilmcor}
\end{table}

\end{document}